# An Automated Multi-Web Platform Voting Framework to Predict Misleading Information Proliferated during COVID-19 Outbreak using Ensemble Method

Deepika Varshney[1], Dinesh Kumar Vishwakarma[2]
Biometric Research Laboratory, Department of Information Technology,
Delhi Technological University, Delhi-110042, India
Email: deepikavarshney06@gmail.com[1], dinesh@dtu.ac.in[2]

**Abstract**

Spreading of misleading information on social web platforms has fuelled huge panic and confusion among the public regarding the Corona disease, the detection of which is of paramount importance. To address this issue, in this paper, we have developed an automated system that can collect and validate the fact from multi web-platform to decide the credibility of the content. To identify the credibility of the posted claim, probable instances/clues(titles) of news information are first gathered from various web platforms. Later, the crucial set of features is retrieved that further feeds into the ensemble-based machine learning model to classify the news as misleading or real. The four sets of features based on the content, linguistics/semantic cues, similarity, and sentiments gathered from web-platforms and voting are applied to validate the news. Finally, the combined voting decides the support given to a specific claim. In addition to the validation part, a unique source platform is designed for collecting data/facts from three web platforms (Twitter, Facebook, Google) based on certain queries/words. This unique platform can also help researchers build datasets and gather useful/efficient clues from various web platforms. It has been observed that our proposed intelligent strategy gives promising results and quite effective in predicting misleading information. The proposed work provides practical implications for the policy makers and health practitioners that could be useful in protecting the world from misleading information proliferation during this pandemic.

## 1. Introduction

In the recent scenario, a new coronavirus disease spread around the world. The disease emerges as a respiratory infection with significant concern for global public health hazards.

Initially, it is suspected that the disease is transmitted from animal to humans, later the paradigm is shifted that the infection is transmitted towards human to human via droplets, close contacts creating huge panic with approximate 6,359,182 confirmed cases and 380,663 deaths[1] have been encountered till now and growth rate is still high which has alarmed the global authorities including world health organization (WHO) [1]. The COVID-19 pandemic affects worldwide badly, however, there is no shortage of people who are taking this crisis as an opportunity for malicious activities/gaining profit[2]. A lot of health-related misleading information, some of the fake cures are suggested for COVID-19 has been posted by the malicious users and creates lots of confusion and misconceptions about the disease. During this pandemic, people have their eye on any new announcement from the government official or some news that can help to get rid of COVID-19. As the disease is deadly, the people are also desperate to know some cure and in rush to find a treatment for the new coronavirus disease. Some of the fake cures posted over social media are harmful too and give bad health advice. The recent examples of fake cures are shown in Fig.1, where Fig.1(a) shows an image with a false claim is gone viral that drinking water a lot and gargling with warm water and salt or vinegar eliminates the coronavirus, however, there is no significant evidence has been found concerning to this claim. Another cure in Fig1(b) reported that the silver solution can kill coronavirus within 12 hours. The proliferation of these misleading information creates lots of misconceptions in the mind-set of people related to coronavirus disease and some of the users are spreading it without verification and fuelled panic among people regarding the COVID-19. According to [3], misleading or fake can be defined as any post that shares content that does not faithfully represent the event that it refers to. We followed this definition in our work and define *"misleading information as the content that does not faithfully represent the event that it refers to and having no significant evidence of proof to validate the claim"*. From the recent research, it has been observed that numerous misleading content is circulating about the coronavirus and it is becoming difficult to differentiate fake news from the real one [4]. The propagation of misleading content on the virus could also be deleterious to mankind. This has led to the dire need for a system that can differentiate fake from real. Earlier many of the previous research has been reported methods of detecting fake news in online social media considering a variety of applications[5]. Most of the previous research has counter fake news problems mainly in the following type: Image-based algorithm and Text-based algorithms[6][7]. Many previous studies have worked on fake news detection by applying a text-based approach. Text-based approaches mainly use text patterns and match them with already existing patterns of fake news. They are sometimes referred to as the linguistic approach. Along with this lots of researchers have shifted their interest in the credibility detection of posts/tweets using text-based features[8][9]. Like a text-based approach, research has also been done by employing an image-based approach. From the study, it has been seen that researchers have explored images based algorithms for the analysis of fake images or images attached with false claims in mainly following ways, Text additive images, and Manipulated images. The manipulated images termed as an image whose piece/part or certain region is manipulated with respect to visual context. Various image-based features have been explored for the classification of images. The authors of [10], propose 5 visual features and 7 statistical features for the verification of news events. Along with the manipulated images, some of the researchers have also considered text-additive images for the analysis of misleading content. The text additive images termed as images embedded with false claims instead of having any manipulation from visual context. The authors of [11], have incorporated text additive images, where they have applied a rule-based algorithm for the prediction of fake news. From recent research, it has been observed that none of the works have shown and reported fake news prediction analysis propagated during one of the major pandemic "CORONAVIRUS". Many people are sharing fake cures to get rid of coronavirus disease without any verification and create lots of misconceptions. Government and officials have also urged peoples to check the authenticity of the post before sharing [4]. This also motivates us to build an intelligent system for the prediction of fake news spreading during this pandemic. We, therefore, developed a generalized multi web platform framework of detecting misleading content on social media platforms, where we have considered COVID-19 as a special issue which is a huge pandemic and taken as one of the application case studies in this work. However, our model is generalized and works for other applications as well. COVID-19 is an emerging issue and very few research have been reported yet in this context that leads to motivates us to build an efficient framework to predict misleading content spreading during the COVID outbreak. The major key contributions of the work are highlighted in the following points.

---

[1] https://in.search.yahoo.com/search?fr=mcafee&type=E211IN885G0&p=coronavirus

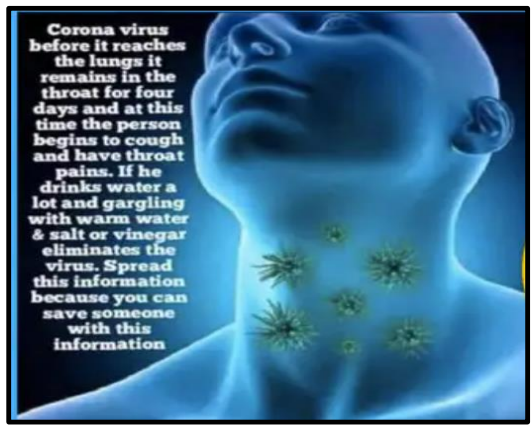 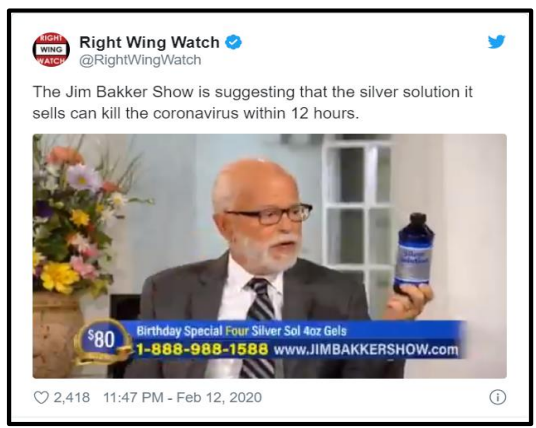

(a)           (b)

Fig 1. Examples of Misinformation[1]

- The proposed work gives a significant contribution in providing a novel generalized Automated Multi-Web Platform Voting Framework for collecting and validating misleading content in an online social network where considering COVID-19(fake news spreading during Corona outbreak) one of the special case studies from the application perspective.
- To the best of our knowledge, we are first to build a unique platform (Facts collector) for the collection of crucial facts and knowledge from three different prominently used social media and web search platform (Twitter, YouTube and Google) as well as provide different mechanism to search the query (build query) to get efficient and relevant results.
- The four set of novel features based on content, linguistics/semantic cues, similarity and sentiments has been gathered from web-platforms that further fed into ensemble based machine learning model to classify the news as Misleading or real. Finally, voting is applied to validate the news and to check the confidence/support given by different web platforms.
- As the COVID-19 is one of the emerging issues and none of the work has been reported yet to predict the fake news propagating during this phase and gives a major contribution by providing the analysis, which greatly helps researchers for further study.
- We investigate the model performances with different classifiers, and comparative analysis reveals that our proposed method outperforms other states of the art on the same dataset.

The remainder of this paper is organized as follows. In Section 2, we are going to discuss the previous work that has been done related to this field, wherein Section 3, we discuss the problem statement and unique fact collector platform. In contrast, Section 4, elaborates the strategy/method that we have employed for the misleading information detection, which is followed by a discussion of experimental results in Section 5. Lastly, the paper is concluded with some suggested future work aspects.

## 2. Related Work

In the current era, spreading false information is one of the crucial problems nowadays, where it is quite difficult for the online user to discriminate fake news from the real one and that's why the development of an intelligent system is required. Most of the methods proposed in earlier states-of-the-art[12] to detect misleading information considered it a classification problem intending to associate label as true or false with a particular claim/post. From the survey analysis, it has been observed that the classification approaches are turn divided into approaches based on machine learning and deep learning. A detailed description is given below.

2.1 Machine Learning

From the study, it has been proven that machine learning algorithms are extremely useful in countering numerous problems in the information engineering field. In particular, many of the machine learning approaches implemented for misleading/ fake information detection applied as a supervised learning strategy[13]. In machine learning classification algorithms support vector machines(SVMs) are one of the widely used methods for classifications. The authors of [14], have proposed a method where they employed a graph-kernel-based SVM classifier to detect rumors using propagation structure and content features with an accuracy of 0.91 on Sina-Weibo dataset. Whereas, in [15] the author reported a set of features to distinguish among fake news, real news, and satire. The SVMs are also employed for clickbait detection in [16]. Like SVM, the random forest has also been exploited in numerous works for fake news and rumor detection. Most of the studies have reported random forests as a strong performer among other machine learning algorithms [17], [18], [19], [20], [21]. In [18], the author has proposed a set of temporal, structural, and linguistic features for the classification of rumors in a tweet graph by employing a random forest with an accuracy of 0.90. The random forest has also been used for stance detection in [17],[22]. The comparative studies of a different approach in the context of rumor and fake news have shown competitive performance for logistic regression [19],[23],[24],[25]]. The authors of [26], employed logistic regression for stance classification of news articles or headlines and claims. Another widely studied family of the algorithm proposed particularly for misleading content detection is a decision tree [27]. The effectiveness of decision tree algorithms like j48 with respect to other machine learning paradigms including SVMs has been reported in [[8],[19]]. The authors of [8] have used the content and context-based features to perform credibility evaluation of tweets and the model is performing well with an accuracy of 0.86. In [28], to evaluate the trustworthiness of users in social media via decision tree, the author has proposed a series of user trust metrics and reported an accuracy of 0.75.

2.2  Deep Learning

Deep learning is one of the prominent and widely explored research topics in machine learning. The main advantage of deep learning over traditional machine learning approaches is they are not based on manually crafted features and lead to reduce feature extraction time. Along with this, the deep learning framework can learn hidden representations from simpler inputs both in context and content variation [29]. The two prominent and widely used paradigms in Morden artificial neural network are RNN and CNN. In [29], authors have proposed a novel RNN architectures, namely tanh-RNN, LSTM, and Gated Recurrent Unit(GRU) for the detection of rumors. From the results, it has been found that GRU has obtained the best results in both the datasets considered with 0.88 and 0.91 accuracies, respectively. Whereas, in [30], the author has proposed a multi-task learning approach and designed a multi-task learning framework with an LSTM layer shared among all tasks to counter the problem of rumor classification. Like RNN, CNNs have also been explored and widely studied for image recognition and many other fields of computer vision. However, it now gaining popularity in the NLP field as well [31]. The authors of [22], have explored a technique using CNN with single and multi-word embedding to counter problems concerning both stance and veracity classification of tweets. The author has reported accuracy of 0.70 for stance classification problem and 0.53 for the problem of veracity classification. Whereas, Paragraph embedding is explored to learn the representation of a small group of posts in a specific event and used them as input for their CNN model in [32] and achieved an accuracy of 0.93 for Sina Weibo and 0.77 for Twitter. From the study, it has been observed that most of the recent work has explored the combination of RNN and CNN in their model [[33],[34],[35]]. The authors of [35] proposed an architecture applied on the LIAR dataset that encodes text information via a CNN and metadata about the author of the text using an LSTM layer as well as it has been also found that the hybrid model has proved to outperform all other baselines along with a bi-LSTM architecture with an accuracy of 0.27 on the testing dataset. Whereas, in [[33]] author has proposed an approach based on repost sequence patterns for the detection of false rumours.

All the different approaches discussed above have considered different machine and deep learning methods for the prediction of misleading content. In our study, we have used a multi web platform framework to gather effective clues for predicting false information. The clues can be detected from multiple web platforms to get the strong support as it may happen that one platform may not give effective clues to predict some information but other can. Moving to this concept, instead of relying only on a specific platform for getting information, our proposed model incorporates multi web platform for retrieving clues concerning to specific query. To the best of our knowledge, none of the previous studies has been explored this concept. Along with this, very few studies incorporate the concept of a unique platform that can collect information from various social media and web sources for building data. These all points we will discuss in later sections in detail.

## 3. Problem Definition and Unique Fact Collector Platform

In this paper, we have considered a binary class classification problem. We assume that the posted source claim $c = \{c1, c2, c3 \ldots ck\}$ can be divided into two classes $Class = \{M, R\}$: 1) Real (R), namely the posted claim faithfully represents the event that it refers to, 2) Misleading (M), namely the posted claim does not faithfully represent the event that it refers to. The claims that have been included here are related to coronavirus COVID-19. Take an example of the post related to coronavirus. The post that "chlorine and alcohol products cannot kill viruses within the body" is true information and belongs to the real class since much authoritative and authentic news media have reported relevant news and has also acknowledged by WHO[2]. While one of the posts says that "coronavirus is caused by 5G technology" is absolutely a false rumour, it lacks factual support and deviates from the scientific principles. So, the goal is to learn a classifier from the labelled feature set, that is $f : X^k \mapsto Y^k$ , where $Y^k$ takes one of the two fine-grained classes: {R, M}. Given the input feature set $X^k$, the classifier $f$ can output the classification result for the posted claim $C^k$. In the coronavirus outbreak, lots of misleading information was posted by users, and most of the misleading information reading fake cures, lockdown, and others have gone viral during the period. The posts have gone viral on different social media platforms like Twitter, Facebook, and others. For analysing post from different social media platform or building dataset, we build a Unique Fact Collector Platform that can take user input query and search on three prominently used web platforms (Google, Twitter and YouTube) simultaneously to collect relevant knowledge concerning to query. From the analysis, it has been found that twitter is not taking long queries as an input that's why it can only collect information by putting keyword based search. Whereas, Google web search and YouTube can take long queries as an input. Our Unique Fact Collector Platform allow users to fetch information from all three platforms concerning to a query. The Fig.2. shows the flow diagram of our collector framework.

---

[2] https://www.indiatoday.in/world/story/drinking-alcohol-will-not-protect-you-from-covid-19-says-who-1653555-2020-03-08

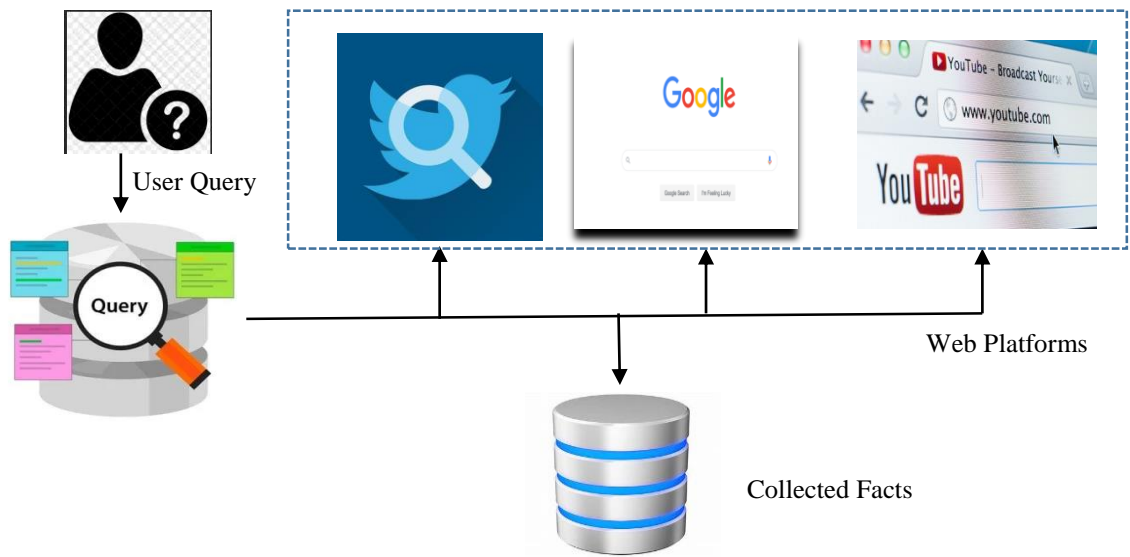

Fig.2. Flow Diagram of Unique Fact Collector Platform

## 4. Proposed Methodology

In this section, we elaborate our automated Multi-Web Platform Voting Framework to Predict Misleading Information Proliferated during COVID-19 outbreak. The detailed flow-diagram of our proposed model is shown in Fig.3. In the first phase of the process the input query is given by the user that he/she wants to validate. The input query is passed through the text processing phase where the cleaning of data has been performed to make it in a format so that it can be used for further processing, and it includes removal of stopwords, removing duplicates, handling missing data, stemming, punctuation removal, text translation (Google translation API) to English language, Removing URLs, symbol, emoji etc. After, text processing the cleaned data is pass to the next module called as "Fact Collection". In this phase, the input is passes through Multi-Web platform to reterive relevant facts concerning to query. The two prominent social media platform that has been utilized to retrieve the facts are

YouTube platform and google web search. In order to gather efficient and relevant titles the query building is one of the important aspects. What query should be pass to get more relevant responses. We have defined three novel way to build a query, however from all the given build cases we have adopted the one which is most effective in reterivng relevant information where we have adopted case 1 as other have some limilation as disccused in Table 1. The Table 1 describes 3 possible cases we have considered for building query.

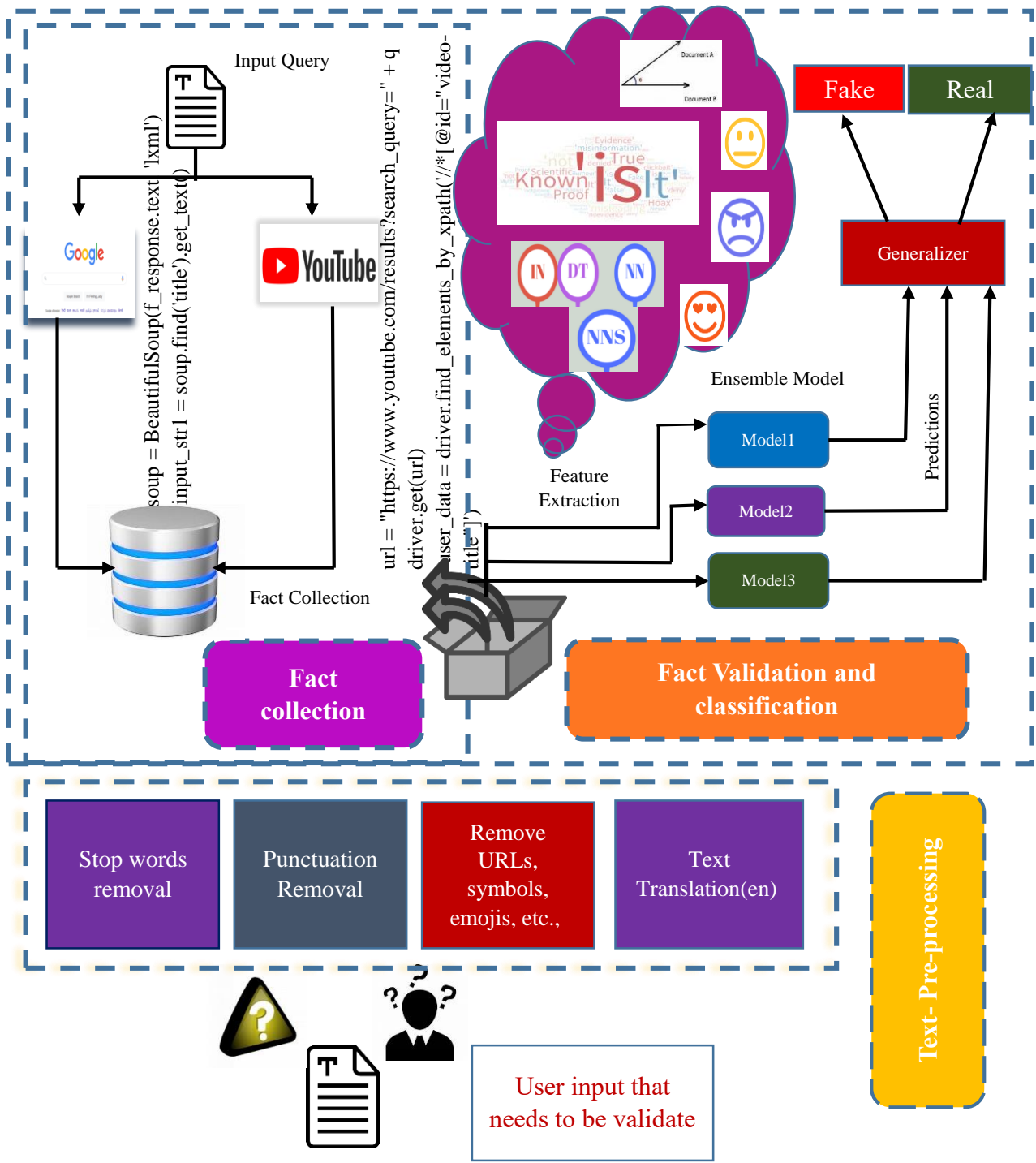

Fig.3. The flow diagram of the proposed approach

Table 1 Possible build cases

| Case No. | Possible Cases | Limitation |
|---|---|---|
| 1. | Stopword_removal(Input_query)+ " "+ "fake news" | - |
| 2. | N_grams(Input_query)+ " " + fake news | 1. Sometimes the context of the query cannot be come out properly and missed out. |
| 3. | Pos(Input_query)+" "+ Fake news | 1. Sometimes the context of the query cannot be come out proprely. <br> 2. In some of the cases giving too many irrelevant facts, goes out of context. |

## 4.1 Fact Collection

In the second phase, the query has been created by removing stopwords to make the query clear and short. As in many cases, the stopwords makes query too long that leads to give irrelevant responses. After removing stopwords, the query is attached with a space concatenated with "*fake news*" keyword. From the analysis, this case of query generation found to be good and considered in this study. Whereas, the other cases includes the N_grams concatenated with the keyword "fake news" and the POS part of speech tagging(proper nouns), we can find all the proper nouns from the input query. Each proper noun is concatenated with the fake news keyword to build a query. However, these case have certain limitations. Sometimes the context of the query cannot become out properly and missed out the relevance.

After this step, the build query is passed through two prominently used platforms, YouTube and Google, to retrieve relevant facts. Finally, the top 10 title headings are scrapped automatically using selenium from both the platform that further be used for analysis. The algorithm for fact collection is shown below in Algorithm 1.

Algorithm 1, shows the process of facts collected from the Multi-web platform. Here, in this study, we have incorporated two social media and web search platforms for reterivng efficient facts/title heading concerning a query that further be used in feature engineering and validate the claim as fake/real. However, other platforms like twitter has also been explored for the collection of facts, but the issue with twitter is it support keyword based searching and long query-based search is not applicable that leads to be a major issue in the collection of relevant facts. Whereas, in the case of google web search and YouTube, we can fetch efficient responses concerning a query.

| Algorithm 1.(Facts Collection) |
|---|
| Input(Text/Text additive image)  and  Output(Collection of Facts) |
|  def. main (): |
|   if(image) |
|      claim= OCR(image) |
|   else: |
|   claim= raw input(claim) |
|   claim= text_preprocessing(claim) |
|   build query= Stopword_removal(claim) + " "+ "fake news" |
|   facts google= google_fact_collect (build query) |
|   facts_youtube= youtube_fact_collect (build query) |
|  def. google_fact_collect(query): |
|     for j in search (query, tld="co.in", num=10, stop=10, pause=2): |
|       f_response= s.get (j, headers = {'User-agent': 'your bot 0.1'}) |
|       soup = BeautifulSoup (f_response. text, 'lxml') |
|       input_str1 = soup. Find('title'). get_text () |
|       input_str1= input_str1.lower() |
|       input_str1= clean(input_str1) |
|       return(input_str1) |
|  def. YouTube_fact_collector(query): |
|     q= query +" "+'fake news' |
|     print(q) |
|     url = "https://www.youtube.com/results?search_query=" + q |
|     print(url) |
|     count=0 |
|     driver.get(url) |
|     user data = driver. find_elements_by_xpath('//*[@id="video-title"]') |
|     print(user_data) |
|     links = [] |
|     titles= [] |
|     for i in user data: |
|      links. append(i.get_attribute('href')) |
|      titles. append(i.get_attribute('title')) |
|      for x in titles: |
|        v_title = x |
|        return(v_title) |

## 4.2  Fact Validation

The third module is fact validation, this module take the facts collected from the previous step and utilizes them to get some efficient clues to predict the claim as fake/real. The four set of features are employed based on content, linguistics/semantic cues, similarity and sentiments. Each of these feature category has been discussed in detail in the below section

Table 2: Detailed description of proposed features

| Features Category | Features | Feature Description |
|---|---|---|
| Content-Based | 1. Question mark count<br>2. Fake word count | 1. Number of question mark in a title heading.<br>2. Number of fake words encountered in a title heading. |
| Linguistics/Semantic cues-Based | 1. NLTK POS TAGGING Semantic Similarity | 1. The nltk wordnet's synset is used to measure the semantic similarity between user query and title headings. |
| Similarity-Based | 1. Cosine Similarity<br><br>$$COS(x, y) = \frac{x.y}{|x|.|y|}$$ | 1. The cosine similarity is used to measure the similarity between user query and title headings. |
| Sentiment-Based | 1. Query sentiments<br>2. Clue sentiments<br>3. Sentiment match count | 1. This features return the sentiment of the user query, either positive negative or neutral.<br>2. This features return the sentiment of the title heading, either positive negative or neutral.<br>3. This features return the count of how many times the query sentiments matchs with the title heading sentiments. |

4.2.1 Content based features:

The content based features has been widely explored in numerous data mining research field. In this paper we have incorporated content based features for the prediction of misleading information including question mark count and fake word count. The question mark count gives necessary clues regarding the confidence reflected from the sentence. If the sentence showing uncertainty, it means that the claim is not sure regarding that event. Question mark count plays a major role in finding the uncertainty in a given sentence. It returns true if any question mark has been encountered in a title/headings retrieved from the web platform, while searching a specific query. Whereas, fake_word_count is also one of the important feature that discriminate fake from real. There are set of false_phrase_corpus that incorporates list of keywords that prominently used to represent fake news. The keyword corpus that we have created including following phrases: { *'false','misleading','inaccurate','rumor','rumour', 'not correct' ,'fake news', 'incorrect', 'wrong', 'confounding', 'deceiving', 'deluding', 'wont', 'did' 'Did', 'funny', 'memes', 'catchy', 'bogus', 'counterfeit', 'fabricated', 'fictitious', 'forged',*

*'fraudulent', 'mock', 'phony', 'affected', 'artificial', 'erroneous', 'fake', 'fanciful', 'faulty', 'improper', 'invalid', 'mistaken', 'unfounded', 'unreal', 'untrue', 'untruthful', 'casuistic', 'fishy', 'illusive', 'imaginary', 'inexact', 'lying', 'misrepresentative', 'falsity', 'misreport', 'misstatement', 'deception', 'falsification', 'artificial', 'fabrication', 'falsehood', 'hoax', '?', 'Not Died', 'misinformation', 'not committed', 'not dead', 'death rumour', 'is it true', 'not known', 'no proof', 'no known', 'no scientific evidence', 'no evidence', 'not verified', 'clickbait', 'not proven', 'denied', 'deny', 'unverified', 'falsely', 'myth', 'ridiculous', 'not true'*}, if any of these word has been encountered in the retrieved responses corresponding to a query, the fake count incremented by 1. The feature is helpful in identifying fake as the title having these phrases more likely representing news as fake.

4.2.2 Linguistics/ semantic cues- based Features

It is very diffcult to process raw text intelligently as the same words in a different order can mean something completely different, while using lingustic knowledge can be possible to solve some problems starting from only the raw characters. For a given claim it is very important to understand in what context it is used. The python library nltk.pos_tag is designed to do the same. When a raw text is passed as an input, it returns a doc object that comes with a variety of annotation. The nltk parse and tag a given document, there are some statistical model which enable it to make prediction of which tag or label most likely applies in this context also called as POS part of speech tagging/ grammatical tagging which is used to mark up a word in a text as corresponding to a particular part of speech, based on both its definition and its context. POS tagging also describes the charactersitics of lexical terms within a sentence or text that further be used for making prediction / assumptions about semantics. To compute the semantic text similarity between two sentence we have used POS (Part of speech) text similarity. There are different POS tags that can be given to a each word in a sentence like $(NNS, noun\ plural)(NNP, proper\ noun, singular)(NN, noun, singular)etc$ .NLTK POS tagger is employed to assign grammatical information of each word of the sentence. This feature is useful in computing the semantic text similarity between the user query and the clues reterived from web platforms. The tags generated by nltk.pos_tag are converted to the tag used by wordnet.synsets. The nltk wordnet's synset is used to measure the similarity.

4.2.3 Similarity based Feature

This is another category of feature used in this work based on similarity. This feature is helpful in segregating relevant title/heading from all the given responses, as not all responses are useful for validation. To get efficient performance of the model we need to remove irrelevant titles from the analysis, only those who cross the threshold value are used for analysis. The one of the prominently used similarity measure "cosine similarity" has been used in this work to compute the similarity between two sentences irrespective of their size. The sentences are considered as two vectors and the cosine similarity between two vectors is measured in 'θ'. If the angle between two sentence is 0 it means they are similar, and if θ = 90° they are dissimilar. The formula of calculating similarity between two sentence x and y can be given as:

$$COS(x,y) = \frac{x.y}{|x|.|y|} \quad (1)$$

4.2.4 Sentiment based:

Sentiment based features are the fourth set of features employed for the prediction of fake news. Sentiments plays an important role in identifying polarity of the sentence, whether it is showing positive negative or neutral sentiments, Here, we have considered 3 features under this category.

1) Query Sentiment: Query Sentiment is a sentiment of the input query given by the user.
2) Title/heading sentiment: This is a sentiment of the responses(title/heading) received as a search result concerning a specific query.
3) Sentiment match count: From all the 10 responses retrieved from the web platforms, how many times the sentiments of the query and the titles are matches. It also represents whether the sentiment pose by the input query is equivalent to the responses received. It also means that both query and heading are posing the same sentiments and presented in the same polarity.

All these above discussed features are briefly shown in Table 2 and the Algorithm 2 elaborate the complete process of fact validation, where the functions to evaluate the four set of features are briefly explained that later be fed to ensemble based classifier for analysing the performance of the model.

```
Algorithm 2.(Facts Validation)
Input(facts)  and  Output(status(fake/real))
def. main ():
   Linguistic features=   Feature_extraction_lingustic(facts)
   Content_feature   =    Feature_extraction_content(facts)
   Sentiments_features  = Feature_extraction_sentiments(facts)
   Similarity_features =  Feature_extraction_similarity(facts)
   Classification_model   =   Ensemble_classifier  (Content_features,  Sentiments_features,  Linguistic_features, Similarity_features)
    return(status)
def. Feature_extraction_lingustic(facts):
```

```
                r. extract_keywords_from_text(input_str1)
                ti=r.get_ranked_phrases () # To get keyword phrases ranked highest to lowest.
                print(ti)
                Y1= listToString(ti)
                s1 = nltk.pos_tag(nltk.word_tokenize(input_str))
                s2 = nltk.pos_tag(nltk.word_tokenize(input_str1))
                Semantic_similarity= similarity (s1, s2)
                return(Semantic_similarity)
    def. Feature_extraction_content(facts):
            r. extract_keywords_from_text(input_str1)
            ti=r.get_ranked_phrases () # To get keyword phrases ranked highest to lowest.
            print(ti)
            Y1= listToString(ti)
            if any (word in Y1 for word in punctuation):
                    pun=pun+1
                    print ("pun count", pun)
            if any (word in Y1 for word in keyword):
                    fake_count=fake_count+1
                    print ("The fake count", fake_count)
    def. Feature_extraction_similarity(facts):
                list1 = word_tokenize(input_str)
                list2 = word_tokenize(input_str1)
                X_set = {w for w in list1 if not w in sw}
                Y_set = {w for w in list2 if not w in sw}
                # form a set containing keywords of both strings
                rvector = X_set.union(Y_set)
                for w in rvector:
                 if w in X_set: l1.append(1) # create a vector
                 else: l1.append(0)
                 if w in Y_set: l2.append(1)
                 else: l2.append(0)
                 c = 0
              # cosine formula
                for i in range(len(rvector)):
                 c+= l1[i]*l2[i]
                cosine = c / float((sum(l1)*sum(l2))**0.5)
                print("The similarity: ", cosine)

    def. Feature_extraction_sentiments(facts):
                query_sentiment= get_tweet_sentiment(query)
                print("The sentiment of the query",query_sentiment)
                title_sentiment= get_tweet_sentiment(input_str1)
                print ("The sentiment of the title", title_sentiment)
                if(query_sentiment==title_sentiment):
                 senti_count=senti_count+1
                 print ("The senti_count", senti_count)
```

## 5. Experimental Analysis and Results

In this section, the experimental analysis is performed on publicly available datasets, different performance measures are adopted (Precision, Recall, F1-score, Accuracy etc.) to measure the effectiveness of the proposed method and lastly presenting the results showing the performance of the proposed model as well as comparative analysis with other State-of-the-art methods. This section covers each of these points in the following subsections.

## 5.1 Constraint-2021 COVID-19 Fake News Detection Dataset

In this paper, we have used the constraint-2021 shared task to detect COVID-19 fake news in English. The dataset is collected from various social media like Twitter, Facebook, Instagram, etc. The main objective of this task is to classify a given social media post into Fake/Real. The dataset collects 10,700 manually annotated social media posts and articles of fake and real news on COVID-19 [36]. The dataset is further split into training validation and test sets in the ratio of 3:1:1 as shown in Table 3.

Table 3. The Constraint -2021 task dataset description

| Split category | Real | Fake | Total |
| --- | --- | --- | --- |
| Training set | 3360 | 3060 | 6420 |
| Validation set | 1120 | 1020 | 2140 |
| Test set | 1120 | 1020 | 2140 |
| Total | 5600 | 5100 | 10700 |

The experimental analysis is performed by employing various machine learning algorithms like Logistic Regression(LR), Support Vector Machine(SVM), Random forest, Ensemble based classification model, etc. We employed this dataset to measure the performance of our model with respect to precision, recall, f1-score and accuracy. As we have incorporated multi-web platforms, so analysis has been performed on both the web platform separately as well hybrid (Google +YouTube) of both. The Results concerning to this experiment is shown Table 4, Table 5 and Table 6 respectively. The first study incorporates the clues retrieved from google platform concerning to specific query. The performance has been analysed majorly on four measures Precision, Recall, F1-Score and Accuracy by employing Random Forest, Support Vector Machine, Logistic- Regression and Ensemble based classification algorithm, where it has been observed that ensemble based Voting classifier (RF, LR, KNN) and (LR, LSVM, CART) performs best and outperforms all other classifier with an F1 Score of 0.989. Whereas, the second best run is given by SVM with an F1-Score of 0.987. Another set of experiment has been performed on the clues retrieved from YouTube platform. It has been observed that LR perform best with an F1-Score of 0.869. However, it can be clearly seen that the clues retrieved from the YouTube alone are not well efficient as compare to Google web search platform. Therefore, the concept of Multi Web platform comes into picture, giving support to a claim in making decision as false or true, if some other platform cannot give sufficient clues for the prediction. The hybrid model improves the performance of the model by incorporating clues concerning to both the web platform to improve the efficiency of the model. The best F1-Score on hybrid model is achieved from SVM and ensemble based classifier (LR, LSVM, CART) having value 0.980 as shown in Table 6. Some earlier studies have also worked on the given problem and reported results concerning the Constraint Task 2021 Covid fake news dataset. The Comparative Study with the other state-of-the-art method on the validation set is shown in Table 7. The authors of [37], proposed a method to predict misleading information proliferated during COVID-19 outbreak, by incorporating various machine learning classifiers and here represented as Model 1. The best run provided by Model 1 by using SVM with an F1-Score of 95.70, whereas our proposed approach on SVM giving the F1-Score of 98.70 and enhanced the performance by 3% as shown in Fig.4. Similarly, the authors of [36] (Model 2) and [38](Model 3) , worked on the same problem task using machine learning and ensemble based classification approach, where they reported best run F1-Score of 93.46 by employing SVM and 98.32 using Ensemble based model respectively as shown in Fig.5. and Fig.6. respectively. It can be clearly that that our model outperforms in all discussed cases and provided the best run using ensemble based model incorporating (LR, LSVM and CART) with an F1-Score of 98.88.

Table 4. Performance of the model incorporating Google web platform

| Model | Precision | Recall | F1-Score | Accuracy |
|---|---|---|---|---|
| Random Forest | 0.986 | 0.986 | 0.986 | 0.985 |
| SVM | 0.987 | 0.987 | 0.987 | 0.987 |
| LR | 0.986 | 0.986 | 0.986 | 0.986 |
| Ensemble learners | | | | |
| Random Forest | 0.986 | 0.986 | 0.986 | 0.985 |
| Voting classifier (RF, LR, KNN) | 0.989 | 0.989 | **0.989** | 0.989 |
| Voting Classifier (LR, LSVM, CART) | 0.989 | 0.989 | 0.989 | 0.987 |
| Bagging Classifier(Decision-Tree) | 0.980 | 0.979 | 0.979 | 0.978 |

Table 5. Performance of the model incorporating YouTube web platform

| Model | Precision | Recall | F1-Score | Accuracy |
|---|---|---|---|---|
| Random Forest | 0.866 | 0.852 | 0.850 | 0.851 |
| SVM | 0.860 | 0.860 | 0.860 | 0.860 |
| LR | 0.870 | 0.869 | 0.869 | 0.869 |
| Ensemble learners | | | | |
| Random Forest | 0.866 | 0.852 | 0.850 | 0.851 |
| Voting classifier (RF, LR, KNN) | 0.863 | 0.863 | 0.863 | 0.862 |
| Voting Classifier (LR, LSVM, CART) | 0.865 | 0.865 | 0.865 | 0.864 |
| Bagging Classifier(Decision-Tree) | 0.853 | 0.795 | 0.785 | 0.795 |

Table 6. Performance of the model incorporating both the Web Platform (Google + YouTube)

| Model | Precision | Recall | F1-Score | Accuracy |
|---|---|---|---|---|
| Random Forest | 0.974 | 0.973 | 0.973 | 0.973 |
| SVM | 0.980 | 0.980 | 0.980 | 0.979 |
| LR | 0.978 | 0.978 | 0.978 | 0.976 |
| SGD | 0.980 | 0.980 | 0.980 | 0.975 |
| Ensemble learners | | | | |
| Random Forest | 0.974 | 0.973 | 0.973 | 0.973 |
| Voting classifier (RF, LR, KNN) | 0.979 | 0.979 | 0.979 | 0.979 |
| Voting Classifier (LR, LSVM, CART) | 0.980 | 0.980 | **0.980** | 0.980 |
| Bagging Classifier(Decision-Tree) | 0.972 | 0.971 | 0.970 | 0.975 |

Table 7 Comparative Study with the other state-of-the-art method on the validation set.

| Author | Model | Precision | Recall | F1-Score | Accuracy |
|---|---|---|---|---|---|
| [37] (Model1) | SVM | 95.71 | 95.70 | 95.70 | 95.70 |
| [37] (Model 1) | LR | 95.43 | 95.42 | 95.42 | 95.42 |
| [37] (Model 1) | RF | 90.98 | 90.79 | 90.80 | 90.79 |
| [37] (Model 1) | NB | 93.33 | 93.32 | 93.31 | 93.32 |
| [37] (Model 1) | MLP | 93.62 | 93.60 | 93.59 | 93.60 |
| [36] (Model 2) | DT | 85.31 | 85.23 | 85.25 | 85.23 |
| [36] (Model 2) | LR | 92.76 | 92.79 | 92.79 | 92.75 |
| [36] (Model 2) | SVM | 93.46 | 93.48 | 93.46 | 93.46 |
| [38] (Model 3) | Ensemble based model | 98.32 | 98.32 | 98.32 | 98.32 |
| Our proposed model | Ensemble voting classifier(LR,CART,LSVM) | 98.88 | 98.88 | **98.88** | 98.79 |
| Our proposed model | Random Forest | 98.20 | 98.10 | 98.10 | 98.09 |
| Our Proposed model | LSVM | 98.70 | 98.70 | 98.70 | 98.70 |
| Our Proposed model | Logistic Regression | 98.60 | 98.60 | 98.60 | 98.55 |
| Our Proposed Approach model | NB | 95.55 | 95.53 | 95.54 | 95.34 |

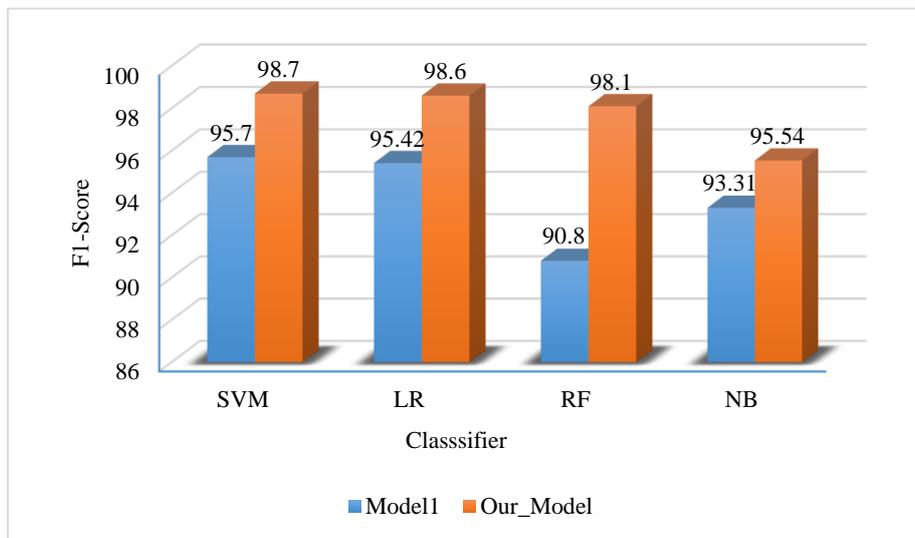

Fig.4. Comparative analysis with Model 1 on F1 score

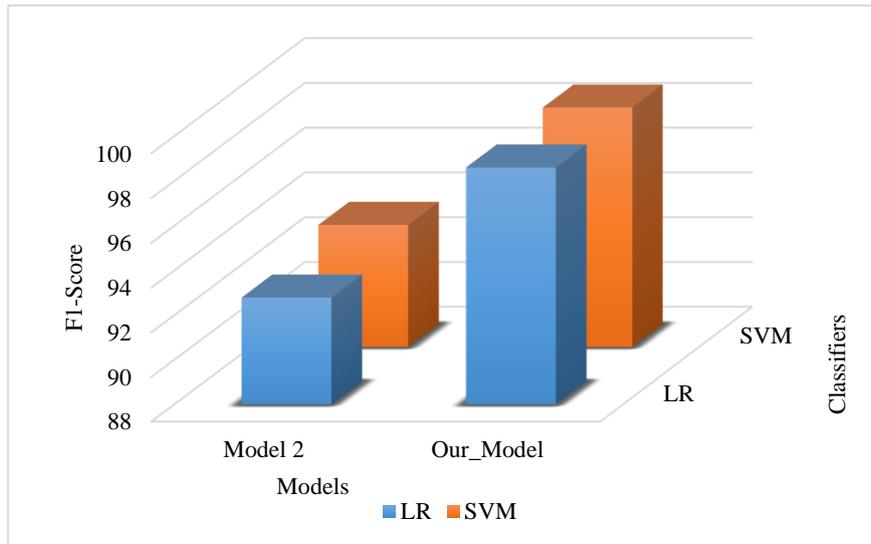

Fig.5. Comparative analysis with Model 2 on F1 score

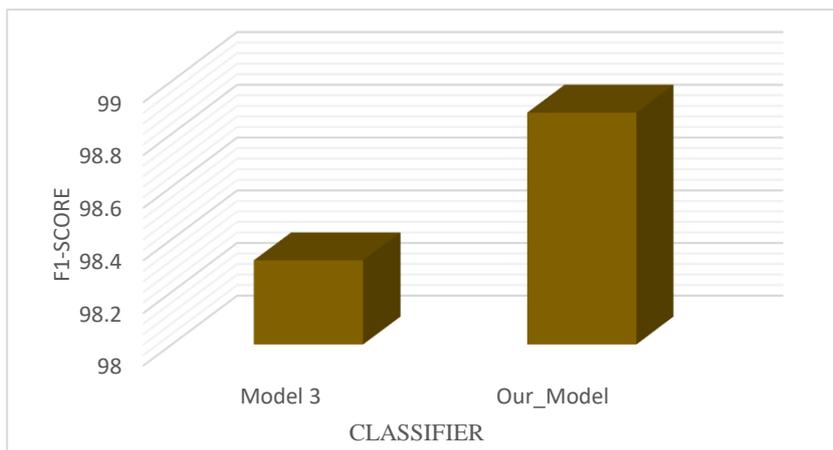

Fig.6. Comparative analysis with Model 3 on F1 score

## 6. Conclusion and Future Work

In this paper, we developed an intelligent generalized strategy for identifying possible clues to predict misleading information, where fake news proliferated during the COVID-19 outbreak is considered as a special case study and detailed analysis has been discussed. We proposed an automated Multi-Web Platform Voting Framework considering YouTube and Google as major sources for the retrieval of clues. The four set of novel features based on content, linguistics/semantic cues, similarity and sentiments has been gathered from these platforms that further fed into ensemble based machine learning model to classify the news as Misleading or real. Voting is applied to validate the news and to check the confidence/support given by different web platforms. It has been observed that Google web platform itself performing good in retrieving crucial knowledge, giving best F1-Score of 98.89 by employing Ensemble based model incorporating LR, LSVM and CART and their voting gives the final decision. However, considering YouTube as a web platform alone for retrieving knowledge it only is able to give an F1-Score of 86.90 by employing LR which is quite low. Here, we can see YouTube alone is not able to retrieve effective clues to predict the news, however, incorporating multi web platform scheme we can improve the performance of the model by taking support from other platforms to validate the veracity of news when it is not available. Retrieving clues from multi-web platform improve the performance of the model and it outperforms other state-of the-art technique on the same dataset by employing ensemble based classification model. In the future we are planning to incorporates and explore other platforms (Instagram, WhatsApp etc.) to validate the news as well as also expand the work by including different modalities of data (images, videos, etc). Along with this, we are also planning to build a real time application for the users to predict misleading content.